# Ambient backscatters-friendly 5G networks: creating hot spots for tags and good spots for readers

Romain Fara
Orange Labs Networks, Châtillon, France.
Laboratoire des Signaux et Systèmes,
University of Paris-Saclay, CNRS,
CentraleSupélec, Gif-Sur-Yvette, France
romain.fara@orange.com

Dinh-Thuy Phan-Huy
Radio Innovation,
Orange Labs Networks,
Châtillon, France
dinhthuy.phanhuy@orange.com

Marco Di Renzo
Laboratoire des Signaux et Systèmes,
University of Paris-Saclay, CNRS,
CentraleSupélec, University of Paris-Sud,
Gif-Sur-Yvette, France
marco.direnzo@l2s.centralesupelec.fr

*Abstract*—In this paper, we present an ambient backscatters-friendly 5G network that creates locations with large power (hot spots) for tags and good reception locations (quiet spots or coherent spots) for readers. The massive multiple input multiple output (M-MIMO) antenna and beamforming capability of the 5G network is used as follows. In a first step, a training device (separate from tags and readers) is used to send pilots and train the 5G network to perform focusing and/or nulling onto marked locations. In a second step, tags and readers are positioned onto the marked locations. The robustness of M-MIMO beamforming to slight changes in the environment is exploited. Our initial simulation results, in a multipath propagation channel environment, show that creating a hot spot on a tag improves the tag-reader range however with a low probability of detection. Creating a hot spot on a tag and a good reception spot on a reader at the same time, improves the tag-reader range with 99% probability of detection. The study also shows that beamforming does not degrade the performance of a legacy 5G communication.

## I. Introduction

Even though each new generation of mobile network, from the $2^{nd}$ generation (2G) to the $5^{th}$ generation (5G) has been designed to be more energy efficient, mobile networks still keep on spending more energy due to the huge wireless internet traffic growth [1], especially due to the internet-of-things (IoT).

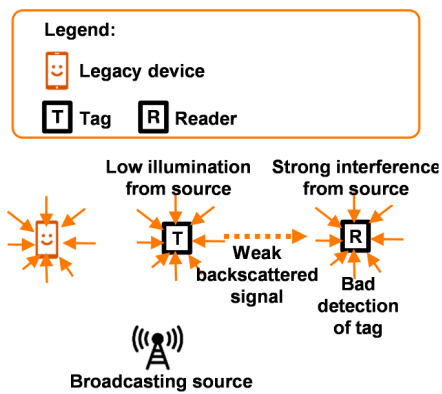

Fig. 1. Illustration of the two limitations of an ambient backscattering communication system in a richly scattering propagation environment: the tag may happen to be in a deep fade of the source and the reader on a strong peak of the source, which may lead to a bad detection of the tag by the reader.

It is therefore time to consider radically different approaches such as ambient backscatters [2]. In an ambient backscatter communication system, a radio-frequency (RF) tag transmits a message to an RF reader, without battery, and without generating any additional RF wave. The tag is simply illuminated by an RF source such as a television broadcast tower, and it switches between two states (in one state it is transparent to the RF signal, in the other state it backscatters the RF signal). One state corresponds to "0" bit and the other to "1" bit. The reader detects two distinct levels of received power associated to the two states, and deduces the sent bit. [3] has shown that ambient backscatter communications are particularly advantageous for IoT applications as the tags are battery-free devices. Unfortunately, the ambient backscatter has two drawbacks, that limit the tag-reader communication range the tag might not be well illuminated and the tag-to-reader link is interfered by the direct source-to-reader link. As illustrated in Fig. 1, due to the random nature of fast fading in a richly scattering environment, this drawback may happen often.

To overcome the aforementioned limitations, [4] proposes to use transmit beamforming at the source side and a joint-detection of the source and the tag signals at the reader side. In [4], it is assumed that the source perfectly knows the three channels: the tag-to-reader, the source-to-tag and the source-to-reader channel. Then, the source computes a beamformer that optimizes the detection of the tag, under a constrain (a target detection probability of the source signal). The limit of this technique is that the reader is highly complex and the acquisitions of the three channel state information at the source side, seems very difficult in practice, especially, with a battery-free tag.

Already with 4G, initial experiments of ambient backscatter communications have been successfully demonstrated [5]. 5G base stations can perform beamforming thanks to massive multiple-input multiple-output (M-MIMO) antennas composed of several dozens, to a few hundreds of antenna elements [6][7][8]. In the field trials conducted in [8], sounding reference signals (SRS) [9] are sent by the mobile device in the uplink direction to enable the base station (BS) to acquire the channel state information at the transmission (CSIT) side, channel reciprocity is exploited to perform beamforming.

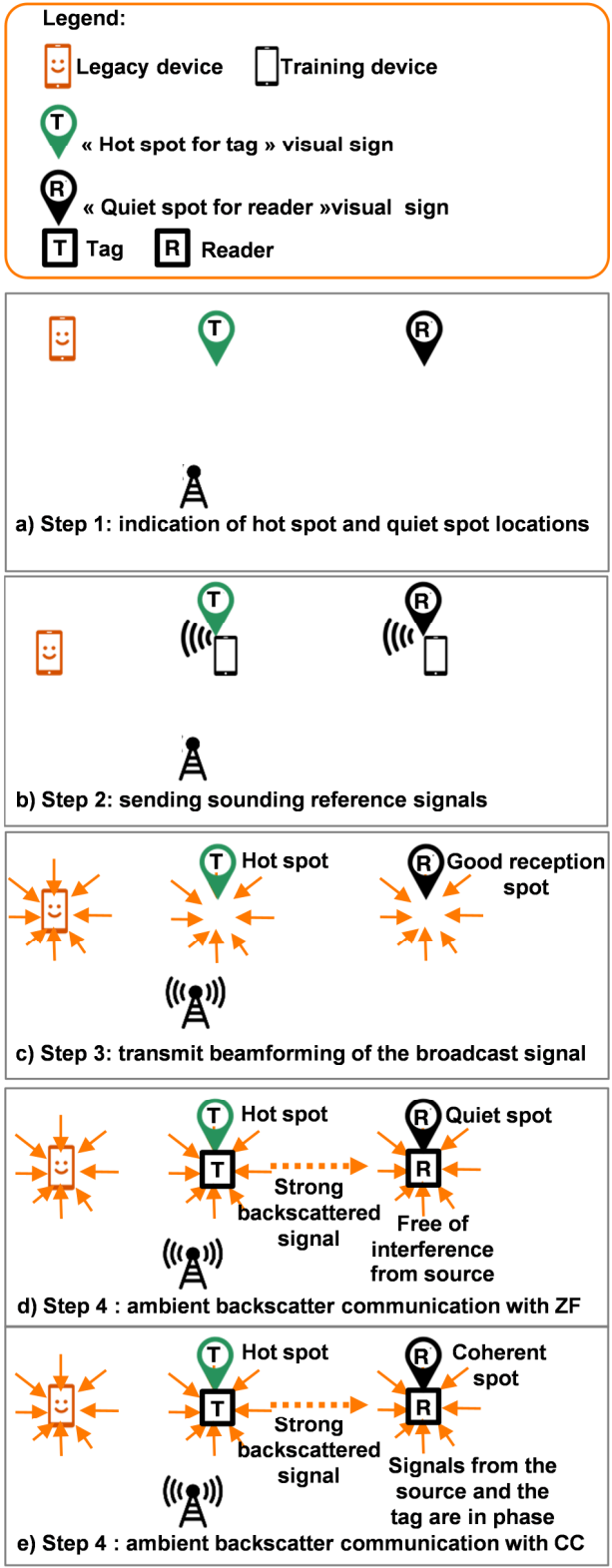

Fig. 2. Propose system

In this paper, for the first time, we propose to use 5G networks and devices to create "hot spots for tags" with very high RF power and "good reception spots for readers" with a four-step approach. We consider two types of good reception spot: a "quiet spot" with very weak source-to-reader interference; and a "coherent spot" with very strong source-to-reader interference, this interference being built to be coherent (i.e. in phase) with the backscattered signal. During a first step illustrated in Fig. 2-a), the locations of the spots are indicated by visual signs such as panels or stickers. During a second step, illustrated in Fig. 2-b), 5G "training devices" send SRS to the network to enable the network to acquire the channel state information at the transmitter side (CSIT). During a third step, illustrated in Fig. 2-c), the network performs maximum ratio transmission (MRT) beamforming [10] (to create a hot spot for the tag) or zero forcing (ZF) beamforming (to create a hot spot for the tag and a quiet spot for the reader simultaneously) [11], or a new coherent combining (CC) beamforming (to create a hot spot for the tag and a coherent spot for the reader). Once the hot spot and good reception spot are created, the 5G devices can be withdrawn and replaced by a tag and a reader, respectively, during a fourth step illustrated in Fig. 2-d) and Fig. 2-e), for ZF and CC, respectively. The tag is then strongly illuminated by the 5G source and better detected by the reader. In the case where the reader is on a quiet spot, the tag-to-reader signal is free of direct 5G source interference. In the case where the reader is on a coherent spot, the tag-to-reader signal coherently combines with the direct 5G source interference. In this latter case the source must transmit successively different phase shifted beams and the reader must analyze the received signal for each beam and feed back the beam index, which maximizes the performance, to the source. Thanks to these methods, the tag and reader remain advantageously simple, low cost and battery-free (for the tag at least), as they do not need to send SRS or to make complex joint detection. The reader can remain a simple energy detector and performance depends on the difference of power between the backscattering and transparent states. Even in the case of CC beamforming, the reader only needs to feed back a beam index.

There are two important requirements though. First, such scheme requires the channel to be stable between the time when the smartphone performs the sounding and the time when the tag and reader communicates with each other. Such stability has been observed experimentally for massive MIMO channels in [10][11]. Secondly, the channel must be highly richly scattering with strong angular. This assumption ensures that applying the ZF, MRT or CC precoder does not reduce the broadcast signal power received by legacy devices, and that the beamforming gain is important.

The paper is organized as follows: section II presents our system model, section III presents simulation results in a richly scattering environment (with a Rayleigh channel model) and section IV concludes this paper.

## II. SYSTEM MODEL

### A. Spatially correlated channel model including backscattering

As shown in Fig. 2, we consider a system composed of a source (S), a tag (T), a reader (R) and a legacy device (D). The variable $d^{SR}$ (respectively $d^{ST}, d^{TR}$) defines the distance between S and R (respectively S and T, T and R).

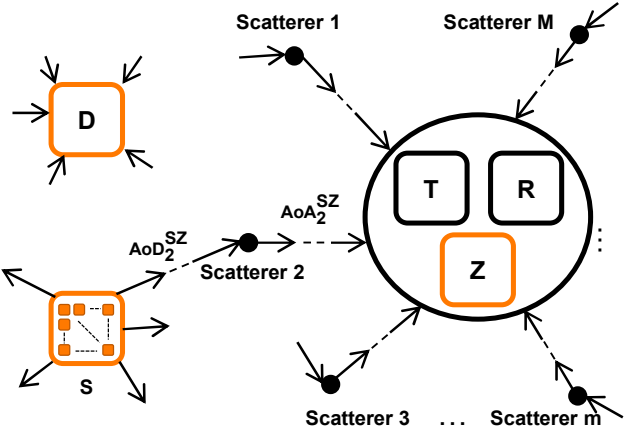

Fig. 3. Propagation model.

The source is equipped with a M-MIMO antenna composed of $K$ transmit antennas. A uniform planar antenna array (with $k^{(1)}$ lines, $k^{(2)}$ columns and $k^{(1)}k^{(2)} = K$) is considered. The source illuminates the tag and the reader through a complex multipath propagation environment due to scatterers.

As illustrated by Fig. 3, we assume that the source is located far from the tag, the reader and the device. In the model we consider that the tag and the reader are closed to each other and experience a spatially correlated channel. On the contrary, we consider that the legacy device is very far from the tag and the reader. As a consequence, the source-to-legacy-device channel is assumed to be uncorrelated with the source-to-reader and source-to-tag channels.

We consider a multi-carrier waveform as in 5G standard [9]. We assume that the channel is frequency flat on a subcarrier and therefore study the system on a subcarrier basis. With this assumption, the channel can be modeled by a complex coefficient. Let $\mathbf{h}^{SZ} \in \mathbb{C}^{1 \times K}$ be the vector of the $K$ coefficients of the small scale fading channels between an arbitrary point in space Z (in the proximity of the tag and the reader) and the $K$ antennas of the source. We model the channel by a Rayleigh channel model with $M$ paths. For each path $m \in [1, ..., M]$, a planar wave leaves the source with a random angle of departure $AoD_m^{ST}$, is scattered by the scatterer $m$, and finally hits the point Z with a random angle of arrival $AoA_m^{ST}$. The wave has a random initial phase $\phi_m$ and a normalized Gaussian complex gain $\alpha_m$. $AoD_m^{ST}$, $AoA_m^{ST}$ and $\phi_m$ are uniformly distributed in $[0, 2\pi]$ and $\mathbb{E}[|\alpha_m|^2] = \frac{1}{M}$. With these assumptions, the expression of $\mathbf{h}_k^{SZ}$ is given by:

$$\mathbf{h}_k^{SZ} = \sum_{m=1}^{M} \alpha_m e^{-j\frac{2\pi f \theta_{k,m}}{c} + \phi_m}, \quad (1)$$

where, $c$ is the light velocity, $f$ is the carrier frequency, $(x_k^S, y_k^S)$ are the Cartesian coordinates of the source antenna $k$, $(x^Z, y^Z)$ are the Cartesian coordinates of the point Z relatively to a position in the area including the tag and reader, and $\theta_{k,m}$ is a path difference given by: $\theta_{k,m} = (x_k^S - x_1^S)\cos(AoD_m^{SZ}) + (y_k^S - y_1^S)\sin(AoD_m^{SZ}) + x^Z \cos(AoA_m^{SZ}) + y^Z \sin(AoA_m^{SZ})$.

From equation (1), we can deduce the vectors $\mathbf{h}^{SR}, \mathbf{h}^{ST} \in \mathbb{C}^{1 \times K}$, of the source-to-reader and the source-to-tag small-scale fadings, respectively, by replacing Z by R and T, respectively.

Free space propagation is considered between T and R and modeled with the Friis Formula as T and the R are supposed to be closed to each other. With these assumptions, the tag-to-reader channel $h^{TR} \in \mathbb{C}$ is given by:

$$h^{TR} = \frac{\lambda}{4\pi d^{TR}} e^{-j\frac{2\pi d^{TR}}{\lambda}}. \quad (2)$$

The total signal received by the reader is the sum of the direct signal from the source and the signal from the source backscattered by tag. We obtain the following expression of the equivalent channel $\mathbf{h}^{eq} \in \mathbb{C}^{1 \times K}$ between S and R (after propagation, including backscattering):

$$\mathbf{h}^{eq} = \sqrt{G}(\gamma h^{TR}\mathbf{h}^{ST} + \mathbf{h}^{SR}) \quad (3)$$

where, $G \in \mathbb{R}^+$ is the is the long-term average propagation gain due to slow shadow fading and the distance, $\gamma$ corresponds to the modulation factor of the tag. $\gamma \in [0,1]$ takes two distinct values: $\gamma^{ON}$ when the tag is backscattering and $\gamma^{OFF}$ when the tag is transparent.

*B. Precoders*

As introduced in I, we propose to use precoders exploiting the knowledge of the channels $\mathbf{h}^{SR}$ and $\mathbf{h}^{ST}$ (estimated by the source thanks to pilots sent by the training device), to improve the $\Delta SNR$ metric. We emphasize that in our proposed schemes, $\mathbf{h}^{TR}$ remains unknown, contrary to the solution given in [4].

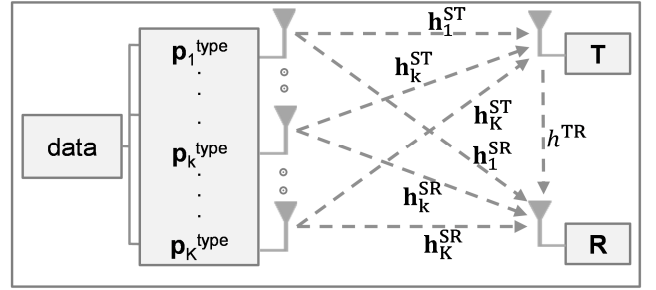

Fig. 4. Precoder

As illustrated in Fig. 4, we consider and compare the following four different types of precoder $\mathbf{p} \in \mathbb{C}^{K \times 1}$:

- the reference precoder $\mathbf{p}^{REF}$ (no precoding);
- the MRT precoder $\mathbf{p}^{MRT}$;
- the ZF precoder $\mathbf{p}^{ZF}$;
- the CC precoder $\mathbf{p}^{CC}$.

The reference (REF) precoder is defined by $\mathbf{p}_1^{REF} = 1$, with $K = 1$ antenna.

For all next precoders, $K > 1$ is considered.

The expression of the MRT precoder is given by $\mathbf{p}^{MRT} = \alpha^{MRT}(\mathbf{h}^{ST})^{\dagger}$, where $(.)^{\dagger}$ is hermitian operation and $\alpha^{MRT}$ is a normalizing factor such that $\|\mathbf{p}^{MRT}\|^2 = 1$, where $\|.\|^2$ is the Frobenius norm. The precoder $\mathbf{p}^{MRT}$ creates a maximum of power or "hot spot" on the tag.

The ZF precoder $\mathbf{p}^{ZF}$ is determined in several steps. First the channel matrix $\mathbf{H} \in \mathbb{C}^{2 \times K}$ is defined as follows:

$$\mathbf{H} = \begin{bmatrix} \mathbf{h}_1^{ST} & ... & \mathbf{h}_K^{ST} \\ \mathbf{h}_1^{SR} & ... & \mathbf{h}_K^{SR} \end{bmatrix}$$

Then, the following zero forcing matrix is computed:

$$\mathbf{Q} = \mathbf{H}^\dagger (\mathbf{H} \mathbf{H}^\dagger)^{-1}$$

Finally, $\mathbf{p}^{ZF}$ is extracted from $\mathbf{Q}$ as follows: $\mathbf{p}_k^{ZF} = \alpha^{ZF} \mathbf{Q}_{k,1}$ for $k = 1 ... K$, with $\alpha^{ZF}$ is a normalizing factor such that $\|\mathbf{p}^{ZF}\|^2 = 1$. The precoder $\mathbf{p}^{ZF}$ creates a maximum of power or "hot spot" on the tag and a zero of power or "quiet spot" on the reader.

The CC precoder $\mathbf{p}^{CC}$ is based on ZF precoder to which we apply a phasing matrix T, a power allocation matrix D and combining matrix S:

$$\mathbf{p}^{CC} = \alpha^{CC} \times \mathbf{Q} \times \mathbf{T} \times \mathbf{D} \times \mathbf{S}$$

With, $\mathbf{T} = \begin{bmatrix} 1 & 0 \\ 0 & e^{j\varphi} \end{bmatrix}$, $\mathbf{D} = \begin{bmatrix} \delta & 0 \\ 0 & \sqrt{1-\delta^2} \end{bmatrix}$ and $\mathbf{S} = \begin{bmatrix} 1 \\ 1 \end{bmatrix}$

$\alpha^{CC}$ is a normalizing such that $\|\mathbf{p}^{CC}\|^2 = 1$. The precoder $\mathbf{p}^{CC}$ creates a hot spot on the tag and another hot spot on the reader. The signal on the reader is phased-shifted in order to combine coherently source and tag signals at the reader location. As the power is shared between the two hot spots, we use $\mathbf{D}$ matrix to optimally allocate this power and maximize the performance.

### C. Performance metrics

The SNR received by the reader is given by:

$$SNR = \frac{|\mathbf{h}^{eq}\mathbf{p}|^2 P_u}{P_{noise}} = |(\gamma h^{TR} \mathbf{h}^{ST} + \mathbf{h}^{SR})\mathbf{p}|^2 SNR^{illum} \quad (4)$$

where $P_u$ is the transmit power of the source, $P_{noise}$ is the receiver noise power at the reader side and $SNR^{illum} = \frac{GP_u}{P_{noise}}$ is the "illumination" SNR. $SNR^{illum}$ reflects the average amount of illumination in the area located around T and R, from the source (independently from precoding and small-scale fading effects).

According to [5], the bit error rate (BER) metric (noted $BER$) of the energy detector depends on the SNR difference $\Delta SNR$ between the state when the tag is backscattering and the state when the tag is transparent as follows:

$$BER = \frac{1}{2} erfc(\Delta SNR), \quad (5)$$

where,

$$\Delta SNR = |SNR^{ON} - SNR^{OFF}| \quad (6)$$

and $SNR^{ON}$ ($SNR^{OFF}$, respectively) is the value of $SNR$, when $\gamma = \gamma^{ON}$ ($\gamma = \gamma^{OFF}$, respectively), i.e. when the tag is backscattering (transparent, respectively). In the rest of the paper we consider $\gamma^{ON} = 1$ and $\gamma^{OFF} = 0$, the expression obtained is (with Re(.) being the real part):

$$\Delta SNR = ||h^{TR}\mathbf{h}^{ST}\mathbf{p}|^2 + 2Re(h^{TR}\mathbf{h}^{ST}\mathbf{p}(\mathbf{h}^{SR}\mathbf{p})^*)|SNR^{illum} \quad (7)$$

Table I details the expressions of $\Delta SNR$ for each beamforming scheme.

TABLE I. BEAMFORMING (BF) SCHEMES AND $\Delta SNR$ EXPRESSIONS

| BF | $\Delta SNR/SNR^{illum}$ expressions | |
|---|---|---|
| Any | $\|h^{TR}\mathbf{h}^{ST}\mathbf{p}\|^2$ Useful tag-to-reader | $+ 2Re(h^{TR}\mathbf{h}^{ST}\mathbf{p}(\mathbf{h}^{SR}\mathbf{p})^*)\|$ + interfering source-to-reader |
| MRT | $\|(\alpha^{MRT})h^{TR}\|\mathbf{h}^{ST}\|^2\|^2$ Strong | $+ 2Re((\alpha^{MRT})^2 h^{TR}\|\mathbf{h}^{ST}\|^2(\mathbf{h}^{SR}(\mathbf{h}^{ST})^\dagger)^*)\|$ + random |
| ZF | $\|\alpha^{ZF}h^{TR}\|^2$ Strong | + 0 + canceled |
| CC | $\|\alpha^{CC}\delta h^{TR}\|^2$ Strong | $+ 2Re\left((\alpha^{CC})^2 \delta h^{TR}(\sqrt{1-\delta^2}e^{j\varphi})^*\right)\|$ + strong and coherent |

In the case of CC, the BS determines the phase-shift $\varphi^{max}$ and the power allocation $\delta^{max}$ that maximise $\Delta SNR$, in two steps. First, the BS transmits successively pilots with different $\mathbf{p}^{CC}$ precoders based on different $\varphi$ and $\delta$ values and the reader measures the corresponding $\Delta SNR$. Second, the reader feeds back the index of the optimal precoder.

To meet a given target quality of service (QoS) associated to a given target BER $BER^{target}$, $\Delta SNR$ must verify: $\Delta SNR > \Delta SNR^{target}$ where $\Delta SNR^{target} = (erfc)^{-1}(2BER)$ is the target SNR difference between the two states.

### D. Metric measuring the impact on legacy device

Regarding the legacy device D, we assume that the legacy D is uncorrelated and far from R and T. The coefficients of the vector $\mathbf{h}^D \in \mathbb{C}^{1 \times K}$ of the source-to-device channel are hence generated as a Rayleigh channel model, independently from $\mathbf{h}^{SR}$ and $\mathbf{h}^{ST}$. We assume that the legacy device D under the average radio conditions (or illumination by the source) as T and R. With these assumptions, the received SNR at the device side is given by:

$$SNR^D = \left|\sqrt{G}\mathbf{h}^D \mathbf{p}^{type}\right|^2 \frac{P_u}{P_{noise}} = |\mathbf{h}^D \mathbf{p}^{type}|^2 SNR^{illum} \quad (8)$$

## III. SIMULATION RESULTS

This section presents simulation results based on the model depicted in II, with the parameter setting given in Table II.

TABLE II. SIMULATION PARAMETERS

| Parameters | Details | Value | Units |
|---|---|---|---|
| $M$ | Number of Taps | 100 | |
| $k^{(1)}$ | Number antenna elements per line | 8 | |
| $k^{(2)}$ | Number antenna elements per colums | 8 | |
| $K$ | Number of antenna elements | 64 | |
| $f$ | Frequency | 2.4 | GHz |
| $SNR^{illum}$ | Average amount of illumination received from the source | [20,30] | dB |
| $BER^{target}$ | Target Bit Error Rate | $10^{-3}$ | |
| $\Delta SNR^{target}$ | Target Difference of SNR | 3.4 | dB |

We first present some visual results for a determined model and then more in depth and general statistical results.

## A. Visualisation of "hot spot" for tag and "quiet spot" for reader thanks to spatial maps of SNR metrics

In this section, we consider fixed locations of the source, the reader and the tag, one sample of the random propagation channel model and a fixed average illumination $SNR^{\text{illum}}$ of 24dB (as defined in section II).

We then compute, the SNR metrics defined in section II, as functions of the (x,y) coordinates of the point Z and visualize them into spatial maps, for each precoder (REF, MRT, ZF and CC, namely) . Fig. 5 illustrates the spatial maps of $SNR^{\text{OFF}}$ and show to which extent the tag and its surrounding area is illuminated by the source. Fig. 6. illustrates the spatial maps of $SNR^{\text{TR}} = |h^{\text{TR}}\mathbf{h}^{\text{ST}}\mathbf{p}^{type}|^2 SNR^{\text{illum}}$ and shows the level of power backscattered by tag. Finally, Fig. 7 illustrates the spatial maps of $\Delta SNR$ maps and shows the locations where the reader can detect the tag with the target QoS (these are locations where $\Delta SNR > \Delta SNR^{target}$). $\Delta SNR^{target}$ is represented as the $\Delta SNR$ level where the scale color changes. Dark colors (blue) indicate locations where QoS cannot be reached and light colors (yellow or red) where QoS can be achieved.

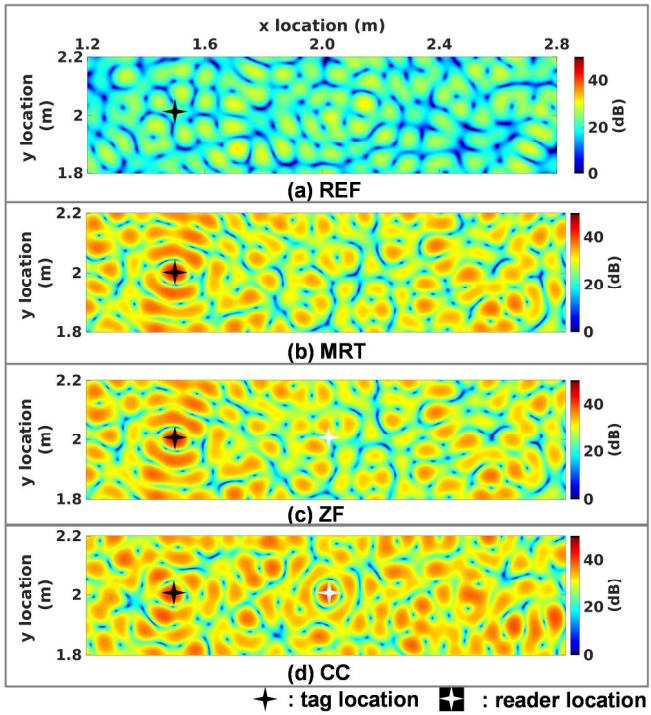

Fig. 5. Maps of $SNR^{\text{OFF}}$.

Thanks to Fig. 5, Fig. 6 and Fig. 7 we can make a first comparison between the four considered precoders. With REF, due to the weak illumination of the tag observed in Fig. 5-a), the backscattered signal observed in Fig. 6-a) is weak, and consequently, the locations observed in Fig. 7-a) where the reader can detect the tag are very rare. With MRT, thanks to the strong illumination (the "hot spot") of the tag observed in Fig. 5-b), the backscattered signal observed in Fig. 6-b) has a large coverage. However, the locations observed in Fig. 7-b), where the reader can detect the tag, are randomly distributed.

At a particular target location of the reader, the target $\Delta SNR$ is not guaranteed.

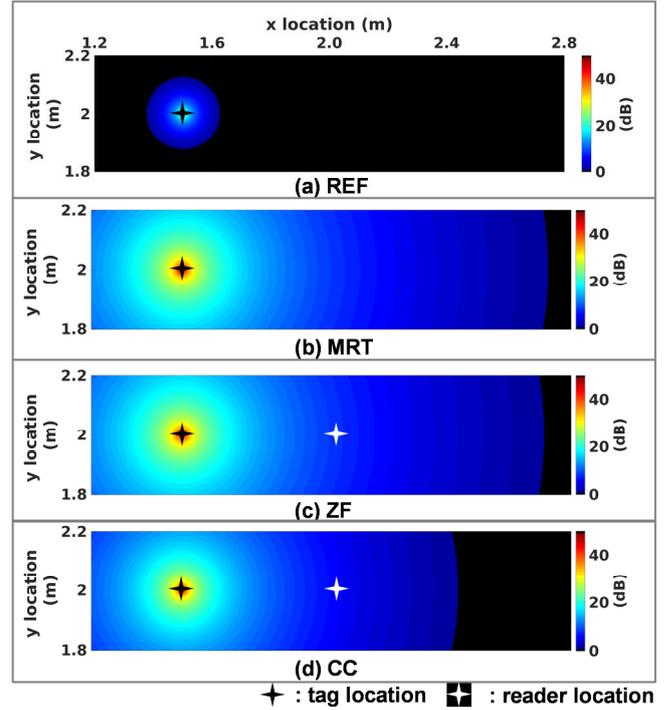

Fig. 6. Maps of $SNR^{\text{TR}}$.

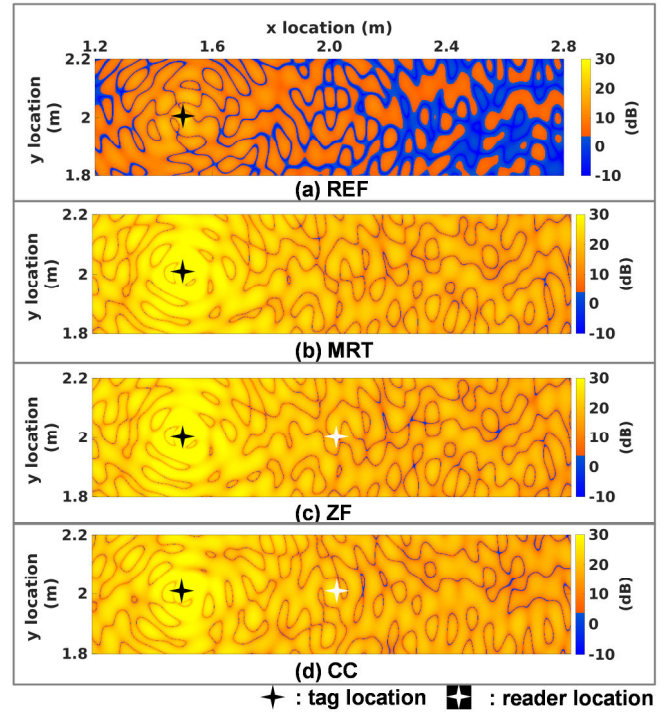

Fig. 7. Maps of $\Delta SNR$.

With ZF, the "hot spot" on the tag observed in Fig. 5-c) is very slightly weaker than in MRT. Hence, the coverage of the backscattered signal observed in Fig. 6-c) is also slightly weaker than in MRT. This is due to the fact that ZF spends some of its power to cancel the source signal on the reader location. However, thanks to this "quiet spot" for reader that can be observed on Fig. 5-c), the target $\Delta SNR$ is met at the target location of the reader as illustrated in Fig. 7-c). Finally with CC, we have plotted the maps for $\varphi = \varphi^{max}$ and $\delta = \delta^{max}$ the phase-shift and the power allocation that maximizes $\Delta SNR$ at the reader location. We observe two hot spots in Fig. 5-d), each of them is weaker than those created by MRT or ZF as the transmit power remains the same and as we have to allocate the power between the tag and the reader. The backscattered signal is consequently weaker as shown in Fig. 6-d). However as the source-reader signal is phased-shifted, the backscattered signal combines with the source signal and improves the detection. We observed in Fig. 7-d) a good detection spot around the reader where we can easily detect the tag.

The previous results have been obtained for a given random sample of the Rayleigh channel model. To get a more general insight on the performance, we propose to visualize statistics over 100 random samples of model parameters. $SNR^{illum}$, the locations of the source and the tag still remain fixed. However, the location of the reader is variable and the ZF and CC beamforming "good reception spot" is adapted to the location of the reader. For each location of the reader, given by the Cartesian coordinates $(x^R, y^R)$, we define $F^O(x^R, y^R)$, as the probability that $\Delta SNR > \Delta SNR^{target}$, i.e. that the reader detects the tag with the target QoS. Fig. 8 illustrates $F^O$ (in percentage) as a function of $(x^R, y^R)$.

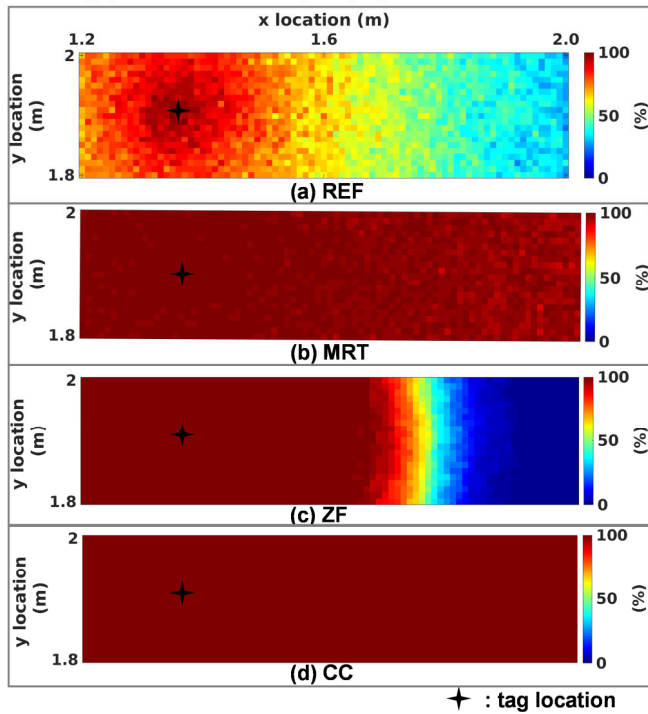

Fig. 8. Statistical maps, evaluating the condition $\Delta SNR > \Delta SNR^{target}$ for 100 random samples of the Rayleigh channel model.

As expected, the reference case is the worst. MRT provides a large coverage, however, coverage holes are randomly distributed. ZF guaranties a continuous coverage inside a given range. CC guaranties a larger continuous coverage thanks to the coherent combining of the backscattered signal with the source signal.

### B. Statistical results as a function of the average illumination $SNR^{illum}$

Previously, the location of the tag and the average illumination $SNR^{illum}$ were fixed and the computation of the metrics was restricted to locations plotted on the maps. In this section, we provide more general results. Statistics of $\Delta SNR$ (over multiple random channel samples) are obtained for various values of $SNR^{illum}$, and reader and tag locations. For each couple of tag and reader locations, the adaptive precoders (MRT, ZF and CC) are updated, i.e. the "hot spot" and "quiet spot" is updated to be focused on the tag and reader, respectively. We define $D^{99\%-REF}(SNR^{illum})$ (resp $D^{99\%-MRT}$, $D^{99\%-ZF}$, $D^{99\%-CC}$)) as the tag-to-reader distance for which, in 99% of the cases, the tag is detected by the reader with the target QoS (i.e. $\Delta SNR > \Delta SNR^{target}$). This distance is expected to increase with average illumination $SNR^{illum}$, as a large $SNR^{illum}$ corresponds to a close or powerful source.

To acquire statistical results of $D^{99\%-REF}$, we draw 20 times the model parameters. We define $SNR^{illum}$ between 20dB and 30dB with a step of 2dB. For each draw of the model and each $SNR^{illum}$ value we define 10 tag locations randomly and uniformly distributed as $x^T \in [0, 100], y^T \in [0, 100]$. For each tag location, reader locations are chosen around the tag as $x^R = x^T + D^{TR} \cos\theta^{TR}$ and $y^R = y^T + D^{TR} \sin\theta^{TR}$, with $\theta^{TR} = \frac{\pi}{10} n^\theta$ ; $n^\theta \in \{0; 1; ...; 19\}$ and $D^{TR} \in \left[\frac{\lambda}{2}, 200\right]$ is chosen to obtain a precision of 1mm. For each configuration of model parameters, $SNR^{illum}$ and locations we evaluate $\Delta SNR$ and the QoS, from these results we can determine the distances $D^{99\%}$ and $D^{90\%}$ for each type of precoder. In the case of CC precoding, we additionally compute the simulation for 360 phase-shift values $\varphi = \frac{2\pi}{360} n^\varphi$ ; $n^\varphi \in \{0; 1; ...; 359\}$ and 10 power allocation $\delta = \frac{2\pi}{10} n^\delta$ ; $n^\delta \in \{0; 1; ...; 9\}$. We evaluate, in the same way, $\Delta SNR$ for every phase-shift and every power allocation, we then conserve the optimal precoder for the performance result and $D^{99\%-CC}$ calculation.

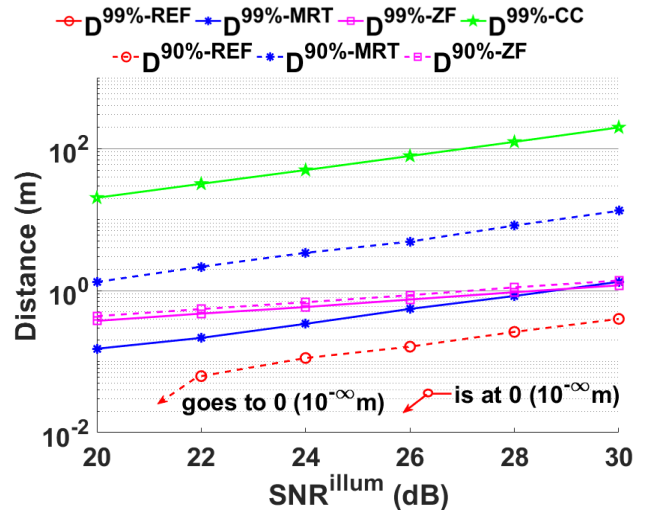

Fig. 9. Distance of 99% and 90% probability detection as function of the average illumination ($SNR^{illum}$).

Simulation has been configured as previously mentioned and statistical results are shown in Fig. 9 that illustrates $D^{99\%-\text{REF}}$, $D^{99\%-\text{MRT}}$, $D^{99\%-\text{ZF}}$, $D^{99\%-\text{CC}}$, $D^{90\%-\text{REF}}$, $D^{90\%-\text{MRT}}$ and $D^{90\%-\text{ZF}}$s as function of $SNR^{\text{illum}}$. As expected, for all precoders, the tag-to-reader distance that guarantees 99% of detection probability increases with the average illumination. For 99% of detection probability, ZF clearly outperforms MRT and REF. However with a lower detection probability target, 90% instead of 99%, MRT outperforms ZF in terms of coverage. As CC guarantees coherent combining of signals, the distance that guarantees 99% of detection probability is far increased.

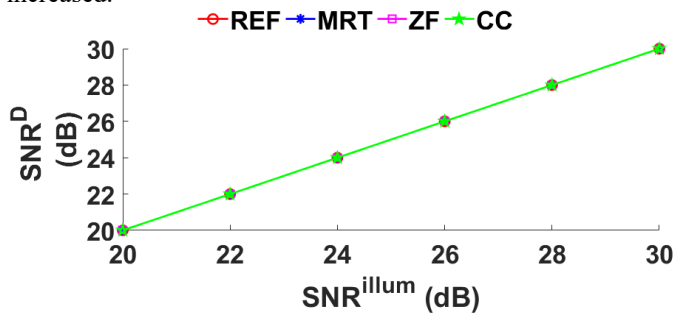

Fig. 10. SNR of the legacy device as fiunction of the average illumination ($SNR^{\text{illum}}$).

Furthermore, Fig. 10 illustrates the received SNR ($SNR^D$) at the legacy device side as a function of the average SNR illumination in four cases: when REF, MRT, ZF and CC precoders are used by the source. No significant difference between the four curves is observed, which means that legacy device is not impacted by the precoder, at least under Rayleigh propagation. In other terms, a backscatter-friendly network that precodes its signals (initially intended towards legacy devices), to help backscatter communications, does not degrade its legacy communications.

## IV. CONCLUSIONS

In this paper, we have shown for the first time a practical method that could be implemented to improve backscatter communications using 5G networks. This method improves the tag-to-reader range and the detection probability. Helped by training devices, the massive MIMO base station acquires a partial knowledge of the channel (only the source-to-tag and the source-to-reader channels) and creates a hot spot to be later used by a tag and a quiet spot to be later used by a reader. We have shown that ZF precoder ensures 99% of detection probability of the tag, with the target quality of service, for short tag-to-reader distances by creating a hot spot on the tag and a quiet spot on the reader. MRT precoder provides larger tag-to-reader distances, if the detection probability target is relaxed and reduced to 90%. CC precoder has advantages of MRT and ZF without the drawbacks. It ensures the QoS for large tag-to-reader distance thanks to the coherent combining of the backscattered signal with the source signal. Moreover we have also shown that precoding has no impact on legacy devices communicating with the 5G network, in Rayleigh environment at least. Next studies will focus on experiments to validate the use of training devices and will explore alternative solutions for radio propagation channels in line-of-sight, and improvements exploiting channel coding.

ACKNOWLEDGMENT

This work is partially supported by the French Project ANR Spatial Modulation under grant ANR-15-CE25-0016 (https://spatialmodulation.eurestools.eu/).

REFERENCES

[1] A. Gati *et. al*., "Key technologies to accelerate the ict green evolution: An operators point of view," in Submitted to IEEE Communications Surveys & Tutorials, available at https://arxiv.org/abs/1903.09627, 2019.

[2] V. Liu, A. Parks., V. Talla, S. Gollakota, D. Wetherall, and J. R. Smith, "Ambient backscatter: Wireless communication out of thin air," in *SIGCOMM 2013,* 2013.

[3] W. Zhang, Y. Qin, W. Zhao, M. Jia, Q. Liu, R. He, B. Ai, "A green paradigm for Internet of Things: Ambient backscatter communications," in *China Communications*, vol. 16, no. 7, pp. 109-119, July 2019.

[4] R. Long, G. Yang, Y. Pei and R. Zhang, "Transmit Beamforming for Cooperative Ambient Backscatter Communication Systems," in *Proc. GLOBECOM 2017 - 2017 IEEE Global Communications Conference*, Singapore, 2017, pp. 1-6.

[5] K. Rachedi, D.-T. Phan-Huy, N. Selmene, A. Ourir, M. Gautier, A. Gati, A. Galindo-Serrano, R. Fara, J. de Rosny "Demo Abstract : Real-Time Ambient Backscatter Demonstration," in *Proc. IEEE INFOCOM 2019*, Paris, France, may 2019.

[6] F. W. Vook, A. Ghosh, E. Diarte and M. Murphy, "5G New Radio: Overview and Performance," in *Proc. 2018 52nd Asilomar Conference on Signals, Systems, and Computers*, Pacific Grove, CA, USA, 2018, pp. 1247-1251.

[7] E. Onggosanusi *et al*., "Modular and High-Resolution Channel State Information and Beam Management for 5G New Radio," in *Proc. in IEEE Communications Magazine*, vol. 56, no. 3, pp. 48-55, March 2018.

[8] T. Kashima *et al*., "Large scale massive MIMO field trial for 5G mobile communications system," in *Proc. 2016 International Symposium on Antennas and Propagation (ISAP)*, Okinawa, 2016, pp. 602-603.

[9] 3GPP TS 38.201 V15.0.0 (2017-12) Technical Specification 3rd Generation Partnership Project; Technical Specification Group Radio Access Network; NR; Physical layer; General description (Release 15).

[10] T. K. Y. Lo, "Maximum ratio transmission," in *IEEE Transactions on Communications*, vol. 47, no. 10, pp. 1458-1461, Oct. 1999.

[11] Sampath, H.; Stoica, P.; Paulraj, A., "Generalized linear precoder and decoder design for MIMO channels using the weighted MMSE criterion," in *Proc. IEEE Transactions on Communications*, vol.49, no.12, pp.2198-2206, Dec 2001.

[12] S. Wesemann, H. Schlesinger, A. Pascht and O. Blume, "Measurement and Characterization of the Temporal Behavior of Fixed Massive MIMO Links," in *Proc. WSA 2017; 21th International ITG Workshop on Smart Antennas*, Berlin, Germany, 2017, pp. 1-8.

[13] D.-T. Phan-Huy, S. Wesemann, J. Björsell and M. Sternad, "Adaptive Massive MIMO for fast moving connected vehicles: It will work with Predictor Antennas! ," in *Proc. WSA 2018; 22nd International ITG Workshop on Smart Antennas*, Bochum, Germany, 2018, pp. 1-8